\documentclass[useAMS]{mn2e}
\usepackage{graphicx}

\begin{document}

\title{Evolution characteristics of central black hole of magnetized accretion disc}

\date{Accepted 2002 May 2.
         Received 2002 April 15; in original form 2002 January 7}

\pubyear{2002} \volume{335} 
\pagerange{655--664} 

\author[D. X. Wang, K. Xiao, W. H. Lei]{D. X. Wang$^*$, K. Xiao, W. H. Lei \\
           Department of Physics, Huazhong University of Science and Technology,Wuhan, 430074, China \\
           $^*$Send offprint requests to: D. X. Wang (dxwang@hust.edu.cn) }

\maketitle
\label{firstpage}

\begin{abstract}
Evolution characteristics of a Kerr black hole (BH) are investigated by considering coexistence 
of disc accretion with the Blandford-Znajek process (the BZ process) and magnetic coupling of 
the BH with the surrounding disc (MC process). (i) The rate of extracting energy from the rotating 
BH in the BZ process and that in MC process are expressed by a unified formula, which is derived 
by using an improved equivalent circuit. (ii) The mapping relation between the angular coordinate 
on the BH horizon and the radial coordinate on the disc is given in the context of general relativity 
and conservation of magnetic flux. (iii) The power and torque in the BZ process are compared with 
those in MC process in detail. (iv) Evolution characteristics of the BH and energy extracting efficiency 
are discussed by using the characteristics functions of BH evolution in the corresponding parameter 
space. (v) Power dissipation on the BH horizon and BH entropy increase are discussed by considering 
the coexistence of the above energy mechanisms.
\end{abstract}

\begin{keywords}
{Accretion,accretion discs -- Black hole physics}
\end{keywords}

\section{Introduction}
{It is well known that magnetized accretion disc of black hole (BH) is an effective model in 
astrophysics. The interaction between a Kerr BH and the surrounding magnetic field has been 
used not only to explain high energy radiation and jet production from quasars and active 
galactic nuclei, but also as a possible central engine for gamma-ray bursts (Rees 1984; Frank, 
King \& Raine 1992; Lee, Wijers \& Brown 2000, hereafter LWB). Confined by the magnetic field 
in the inner region of the disc, the magnetic field lines frozen previously in the disc plasma will 
deposit on the horizon in company with the accretion onto the BH, and the magnetized disc 
becomes a good environment for supporting the magnetic field on the horizon. Blandford and 
Znajek (1977) proposed firstly that the rotating energy and the angular momentum of a BH can 
be extracted by the surrounding magnetic field, and this energy mechanism has been referred to 
as the BZ process, in which the BH horizon and the remote astrophysical load are connected by 
the open magnetic field lines, and energy and angular momentum are extracted from the rotating 
BH and transported to the remote load. 
\\ \indent
There still remain some open problems on the BZ process, and one of them is how to estimate 
the ratio of the angular velocity of the magnetic field lines to that of the BH horizon. Macdonald 
and Thorne (1982 hereafter MT82) argued in a speculative way that the ratio will be regulated to 
about 0.5 by the BZ process itself, which corresponds to the optimal value of the extracting power 
with the impedance matching. However, as argued by Punsly and Coroniti (1990), it is hard to 
understand how the load can conspire with the BH to have the same resistance and satisfy the 
matching condition, since the load is so far from the BH that it cannot be casually connected.
\\ \indent
Recently some authors (Blandford 1999; Li 2000a, 2000b; Li \& Paczynski 2000) argued that with 
the existence of the closed field lines a fast rotating BH will exert a torque on the disc to transfer 
energy and angular momentum from the BH to the disc. Henceforce this energy mechanism is 
referred to as magnetic coupling (MC) process. Compared with the BZ process, the load in MC 
process is the surrounding disc, which is much better understood than the remote load though 
the magnetohydrodynamics (MHD) of the disc is still very complicated.
\\ \indent
Some works have been done to discuss the influence of the BZ process on the evolution of BH 
accretion disc (Park \& Vishniac 1988; Moderski \& Sikora 1996; Lu et al. 1996; Wang et al. 1998, 
hereafter WLY). However MC effects involving the closed field lines were not taken into account 
in the previous works. In fact the closed field lines should exist on the horizon as well as the open 
field lines, and disc accretion should coexist with the BZ and MC process. In this paper evolution 
characteristics of a Kerr BH are investigated by considering coexistence of disc accretion with the 
BZ and MC process (henceforce DABZMC). 
\\ \indent
In order to facilitate the discussion of the evolution of the central BH surrounded by the magnetized 
accretion disc we make the following assumptions: 
\\ \indent
    (i) The disc is perfectly conducting and the magnetic field lines are frozen in the disc. The 
magnetosphere is stationary, axisymmetric and force-free outside the BH and the disc (MT82);
\\ \indent
    (ii) The magnetic field is so weak that its influence on the dynamics of the particles in the disc 
is negligible, and the BH has an external geometry of Kerr metric;
\\ \indent
    (iii) The disc is thin and Keplerian, lies in the equatorial plane of the BH with the inner boundary 
being at the marginally stable orbit.
\\ \indent
This paper is organized as follows. In Sec.II the rate of extracting energy from the rotating BH in the 
BZ process (henceforce the BZ power) and that in MC process (henceforce MC power) are expressed 
by a unified formula, which is derived by using an improved equivalent circuit based on MT82. Our 
result for the BZ power is consistent with that derived in LWB. In Sec.III a mapping relation between 
the angular coordinate on the BH horizon and the radial coordinate on the disc is given in the context 
of general relativity and conservation of magnetic flux. In Sec.IV The power and torque in the BZ process 
are compared with those in MC process in detail. In Sec.V evolution characteristics of the BH and energy 
extracting efficiency are discussed by using the characteristics functions in the corresponding parameter 
space. In Sec.VI we discuss the power dissipation on the BH horizon and the BH entropy increase, and 
show that the excess rate of change of BH entropy does come from the total power dissipation on the BH 
horizon in the BZ and MC process. Finally, in Sec.VII, we summarize our main results.
}

\section{DERIVATION OF BZ POWER AND MC POWER}
{
Two mechanisms of magnetic extraction of energy from a rotating BH are involved in our 
model, i.e., the BZ process and MC process. The configuration of the poloidal magnetic field of 
the BH magnetosphere for the coexistence of DABZMC is shown in Fig.1, where the angular 
coordinate of the open field lines is assumed to vary from $0$ to $\theta_M$, and that of the closed 
field lines from $\theta_M$ to $\pi/2$. $\theta_M$ is a parameter indicating the angular boundary between the 
open and the closed field lines, and the BZ and MC effects vanish as $\theta_M=0$  and $\theta_M=\pi/2$ , 
respectively.

 \begin{figure}
  \resizebox{\hsize}{!}{\includegraphics{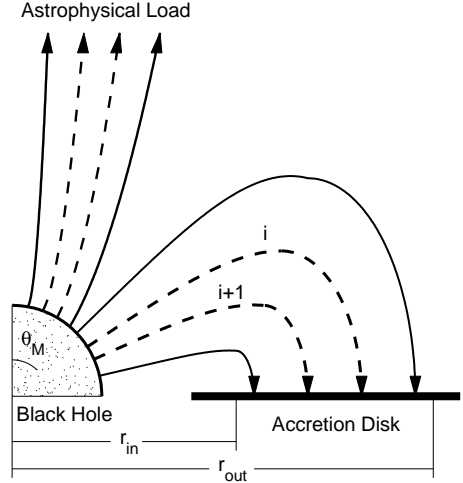}}
  \vspace{-3cm}
  \caption{The configuration of the poloidal magnetic field of the BH magnetosphere.}
  \label{fig:fig1}
\end{figure}

Considering that both mechanisms arise from the rotation of the Kerr BH relative to the 
surrounding magnetic field, we expect that the BZ power and MC power can be expressed by a 
unified formula. In order to do this we proposed a model, in which the remote load in the BZ 
process and the load disc in MC process are incorporated into a unified load with resistance  $Z_L$ 
and angular velocity $\Omega_L$ . Based on MT82 an improved equivalent circuit is shown in Fig.2, in 
which a series of loops correspond to the adjacent magnetic surfaces consisting of the field lines 
connecting the BH horizon and the load.

 \begin{figure}
  \resizebox{\hsize}{!}{\includegraphics{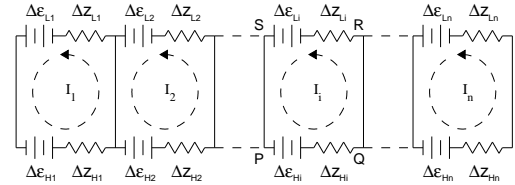}}
  \vspace{-1cm}
  \caption{ An improved equivalent circuit for a unified model for the BZ and MC process.}
  \label{fig:fig2}
\end{figure}

In the i-loop of Fig.2 segments $PS$  (characterized by the flux $\Psi$) and $QR$ 
(characterized by the flux $\Psi+\Delta \Psi$) represent two adjacent magnetic surfaces, and segments $PQ$ 
and $RS$ represent the BH horizon and the load sandwiched by the two surfaces.
$\Delta Z_H$ and $\Delta Z_L$ (hereafter subscript "i" is omitted) are the corresponding resistances 
of the BH horizon and the load, respectively. $\Delta \varepsilon_H=(\Delta \Psi /2\pi) \Omega_H$ and 
$\Delta \varepsilon_L=-(\Delta \Psi/2\pi)\Omega_L$ are electromotive forces due to the rotation of the 
BH and the load, respectively. The minus sign in the expression of $\Delta \varepsilon_L$ arises from 
the direction of the flux.  $\Omega_H$ in the expression of  $\Delta \varepsilon_H$
is the angular velocity of the BH horizon and reads (G=c=1):
\begin{equation}
\Omega _H  = a_ * / ( 2r_H ),r_H = M\left( {1 + q} \right),q = \sqrt {1 - a_ * ^2 } 
\end{equation}
where $a_*$ is the BH spin, which is related to BH mass $M$ and angular momentum $J$ by 
$a_ *   \equiv J / M^2$, and $r_H$ is the radius of the BH horizon. By using the BH magnetosphere model 
given in MT82 we obtain the following equations:
\begin{equation}
\begin{array}{l}
 \int_P^Q {\alpha \vec E \cdot d\vec l}  = \Delta V_H  = \left( {{{\Delta \Psi } \mathord{\left/
 {\vphantom {{\Delta \Psi } {2\pi }}} \right.
 \kern-\nulldelimiterspace} {2\pi }}} \right)\left( {\omega _H  - \Omega _F } \right) \\ 
  = \left( {{{\Delta \Psi } \mathord{\left/
 {\vphantom {{\Delta \Psi } {2\pi }}} \right.
 \kern-\nulldelimiterspace} {2\pi }}} \right)\left( {\Omega _H  - \Omega _F } \right) = I\Delta Z_H  \\ 
 \end{array}
\end{equation}
\begin{equation}
\begin{array}{l}
 \int_R^S {\alpha \left( {\vec E + \vec \upsilon _L  \times \vec B} \right) \cdot d\vec l}  = \Delta V_L  + \left( {{{\Delta \Psi } \mathord{\left/
 {\vphantom {{\Delta \Psi } {2\pi }}} \right.
 \kern-\nulldelimiterspace} {2\pi }}} \right)\left( {\omega _L  - \Omega _L } \right) \\ 
  = \left( {{{\Delta \Psi } \mathord{\left/
 {\vphantom {{\Delta \Psi } {2\pi }}} \right.
 \kern-\nulldelimiterspace} {2\pi }}} \right)\left( {\Omega _F  - \Omega _L } \right) = I\Delta Z_L  \\ 
 \end{array}
\end{equation}
\begin{equation}
\Delta V_L  = \int_R^S {\alpha \vec E \cdot d\vec l}  = \left( {{{\Delta \Psi } \mathord{\left/
 {\vphantom {{\Delta \Psi } {2\pi }}} \right.
 \kern-\nulldelimiterspace} {2\pi }}} \right)\left( {\Omega _F  - \omega _L } \right)
\end{equation}
where $\Delta V_H$ and $\Delta V_L$ are the potential drops measured by "zero-angular-momentum 
observers" (ZAMOs) on the BH horizon and on the load, respectively. ZAMOs is a family of 
fiducial observers defined by Bardeen et al (1972). $\vec \upsilon _L $ is the load's velocity relative to ZAMOs. 
$\omega _H \left( { \to \Omega _H } \right)$ and $\omega_L$ are the ZAMO angular velocities on the BH horizon and the load, 
respectively. $\Omega_F$ is the angular velocity of the magnetic field lines. $\alpha  \equiv \left( {{{d\tau } \mathord{\left/
 {\vphantom {{d\tau } {dt}}} \right. \kern-\nulldelimiterspace} {dt}}} \right)_{ZAMO} $ is the 
"lapse function", which is written as follows:
\begin{equation}
\alpha  = {{\rho \sqrt \Delta  } \mathord{\left/
 {\vphantom {{\rho \sqrt \Delta  } \Sigma }} \right.
 \kern-\nulldelimiterspace} \Sigma }
\end{equation}
where $\Sigma$, $\rho$ and $\Delta$ are the Kerr metric parameters and read
\begin{eqnarray}
\lefteqn{\Sigma ^2  \equiv \left( {r^2  + a^2 } \right)^2  - a^2 \Delta \sin ^2 \theta ,\rho ^2  \equiv r^2  + a^2 \cos ^2 \theta } \nonumber \\  
\lefteqn{\Delta  \equiv r^2  + a^2  - 2Mr }
\end{eqnarray} 
Incorporating equations (2) and (3) we obtain
\begin{eqnarray}
 \Delta \varepsilon _H  + \Delta \varepsilon _L  & = & \int_P^Q {\alpha \vec E \cdot d\vec l}  + \int_R^S {\alpha \left( {\vec E + \vec \upsilon _L  \times \vec B} \right) \cdot d\vec l}  \nonumber \\ 
  & = & I \left( {\Delta Z_H  + \Delta Z_L } \right)
\end{eqnarray}
Dividing equation (2) by equation (3), we derive the relation relating the three angular velocities, 
$\Omega_H$, $\Omega_L$ and $\Omega_F$ as follows:
\begin{equation}
\Omega _F  = {{\left( {\Omega _H \Delta Z_L  + \Omega _L \Delta Z_H } \right)} \mathord{\left/
 {\vphantom {{\left( {\Omega _H \Delta Z_L  + \Omega _L \Delta Z_H } \right)} {\left( {\Delta Z_H  + \Delta Z_L } \right)}}} \right.
 \kern-\nulldelimiterspace} {\left( {\Delta Z_H  + \Delta Z_L } \right)}}
\end{equation}
Assuming $\Omega_L=0$ in the BZ process, we have
\begin{equation}
\Omega _F  = {{\Omega _H \Delta Z_L } \mathord{\left/
 {\vphantom {{\Omega _H \Delta Z_L } {\left( {\Delta Z_H  + \Delta Z_L } \right)}}} \right.
 \kern-\nulldelimiterspace} {\left( {\Delta Z_H  + \Delta Z_L } \right)}}
\end{equation}
Considering that the load disc consists of plasma of perfect conductivity and the magnetic field 
lines are frozen in the disc plasma, we have $\Delta Z_L=0$ and
\begin{equation}
\Omega _F  = \Omega _L  = \Omega _D  = \frac{1}{{M(\chi _{}^3  + a_ *  )}}
\end{equation}
where $\Omega_D$ is the angular velocity of the disc at the place where the magnetic flux penetrates, 
and $\chi=\sqrt{r/M}$ is dimensionless radial coordinate of the disc. The current $I$ in each loop 
can be expressed by
\begin{eqnarray}
 I & = & \frac{{\Delta \varepsilon _H  + \Delta \varepsilon _L }}{{\Delta Z_H  + \Delta Z_L }} 
      = \left( {{{\Delta \Psi } \mathord{\left/ {\vphantom {{\Delta \Psi } {2\pi }}} \right.
         \kern-\nulldelimiterspace} {2\pi }}} \right)\frac{{\Omega _H  - \Omega _L }}{{\Delta Z_H  + \Delta Z_L }} \nonumber \\ 
   & = & \left( {{{\Delta \Psi } \mathord{\left/ {\vphantom {{\Delta \Psi } {2\pi }}} \right.
         \kern-\nulldelimiterspace} {2\pi }}} \right)\frac{{\Omega _H  - \Omega _F }}{{\Delta Z_H }} 
\end{eqnarray}
where equation (8) is used in deriving the right-hand side (RHS) of equation (11). The current 
$I$ on the BH horizon feels Ampere's force, and the torque exerted on the annular ring of the 
width $\Delta l=\rho \Delta \theta$ is
\begin{equation}
\Delta T = \varpi B_H I\Delta l = \left( {{{\Delta \Psi } \mathord{\left/
 {\vphantom {{\Delta \Psi } {2\pi }}} \right.
 \kern-\nulldelimiterspace} {2\pi }}} \right)^2 {{\left( {\Omega _H  - \Omega _F } \right)} \mathord{\left/
 {\vphantom {{\left( {\Omega _H  - \Omega _F } \right)} {\Delta Z_H }}} \right.
 \kern-\nulldelimiterspace} {\Delta Z_H }}
\end{equation}
where $\Delta \Psi  = B_H 2\pi \varpi \Delta l$, $B_H$ is the magnetic field on the horizon, and $\varpi  = \left( {{\Sigma  \mathord{\left/
 {\vphantom {\Sigma  \rho }} \right.
 \kern-\nulldelimiterspace} \rho }} \right)\sin \theta $ is 
the cylindrical radius of the horizon. Therefore the extracting power between the two adjacent 
magnetic surfaces is
\begin{equation}
\Delta P = \Omega _F \Delta T = \left( {{{\Delta \Psi } \mathord{\left/
 {\vphantom {{\Delta \Psi } {2\pi }}} \right.
 \kern-\nulldelimiterspace} {2\pi }}} \right)^2 {{\Omega _F \left( {\Omega _H  - \Omega _F } \right)} \mathord{\left/
 {\vphantom {{\Omega _F \left( {\Omega _H  - \Omega _F } \right)} {\Delta Z_H }}} \right.
 \kern-\nulldelimiterspace} {\Delta Z_H }}
\end{equation}
where
\begin{equation}
\Delta Z_H  = R_H {{\Delta l} \mathord{\left/
 {\vphantom {{\Delta l} {\left( {2\pi \varpi } \right)}}} \right.
 \kern-\nulldelimiterspace} {\left( {2\pi \varpi } \right)}} = {{2\rho \Delta \theta } \mathord{\left/
 {\vphantom {{2\rho \Delta \theta } \varpi }} \right.
 \kern-\nulldelimiterspace} \varpi }
\end{equation}
and $R_H=4 \pi/c=377ohm$ is the surface resistivity of the BH horizon. Considering that the 
BH horizon consists of two hemispheres and integrating equations (13) and (12) over the angular 
coordinate from $\theta=0$ to $\theta_M$ we obtain the total BZ power and torque as follows: 
\begin{equation}
{{P_{BZ} } \mathord{\left/
 {\vphantom {{P_{BZ} } {P_0 }}} \right.
 \kern-\nulldelimiterspace} {P_0 }} = 2a_ * ^2 \int_0^{\theta _M } {\frac{{k\left( {1 - k} \right)\sin ^3 \theta d\theta }}{{2 - \left( {1 - q} \right)\sin ^2 \theta }}} 
\end{equation}
\begin{equation}
{{T_{BZ} } \mathord{\left/
 {\vphantom {{T_{BZ} } {T_0 }}} \right.
 \kern-\nulldelimiterspace} {T_0 }} = 4a_ *  \left( {1 + q} \right)\int_0^{\theta _M } {\frac{{\left( {1 - k} \right)\sin ^3 \theta d\theta }}{{2 - \left( {1 - q} \right)\sin ^2 \theta }}} 
\end{equation}
where $k \equiv {{\Omega _F } \mathord{\left/
 {\vphantom {{\Omega _F } {\Omega _H }}} \right.
 \kern-\nulldelimiterspace} {\Omega _H }}$ and
\begin{equation}
\left\{ \begin{array}{l}
 P_0  = \left< B_H^2 \right> M^2  \approx B_4^2 M_8^2  \times 6.59 \times 10^{44} erg \cdot s^{ - 1}  \\ 
 T_0  = \left< B_H^2 \right> M^3  \approx B_4^2 M_8^3  \times 3.26 \times 10^{47} g \cdot cm^2  \cdot s^{ - 2}  \\ 
 \end{array} \right.
\end{equation}
In equation (17) $\left< B_H^2 \right>$ is the average value of $B_H^2$ over the BH horizon, 
$B_4$ and $M_8$ are $\sqrt{ \left< B_H^2 \right> }$ and $M$ in the units of $10^4 gauss$ and 
$10^8 M_{\sun}$, respectively. Hereafter $\left< B_H^2 \right>$ is regarded as a constant on the 
horizon, and the sign of the average value is omitted. Taking $\theta_M=\pi/2$ and $k=0.5$ in 
equation (15) we obtain the optimal BZ power as follows:
\begin{eqnarray}
\lefteqn{ P_{BZ}^{optimal} / P_0 = Q^{-1} \left( \arctan Q - a_* /2 \right), } \nonumber \\
\lefteqn{ Q \equiv \sqrt{ ( 1-q ) / ( 1+q )} }
\end{eqnarray}
This result is exactly the same as derived by Lee et al, who pointed out that the BZ power had 
been underestimated by a factor ten in previous work (LWB). The corresponding optimal BZ torque 
can be expressed by
\begin{equation}
  T_{BZ}^{optimal} / P_0 = 4Q^{-2} \left( \arctan Q - a_* /2 \right)
\end{equation}
Similarly, taking $\Omega_F=\Omega_D$ and integrating equations (13) and (12) over the angular 
coordinate from $\theta_M$ to $\pi/2$, we obtain total MC power and torque as follow: 
\begin{equation}
{{P_{MC} } \mathord{\left/
 {\vphantom {{P_{MC} } {P_0 }}} \right.
 \kern-\nulldelimiterspace} {P_0 }} = 2a_ * ^2 \int_{\theta _M }^{{\pi  \mathord{\left/
 {\vphantom {\pi  2}} \right.
 \kern-\nulldelimiterspace} 2}} {\frac{{\beta \left( {1 - \beta } \right)\sin ^3 \theta d\theta }}{{2 - \left( {1 - q} \right)\sin ^2 \theta }}} 
\end{equation}
\begin{equation}
{{T_{MC} } \mathord{\left/
 {\vphantom {{T_{MC} } {T_0 }}} \right.
 \kern-\nulldelimiterspace} {T_0 }} = 4a_ *  \left( {1 + q} \right)\int_{\theta _M }^{{\pi  \mathord{\left/
 {\vphantom {\pi  2}} \right.
 \kern-\nulldelimiterspace} 2}} {\frac{{\left( {1 - \beta } \right)\sin ^3 \theta d\theta }}{{2 - \left( {1 - q} \right)\sin ^2 \theta }}} 
\end{equation}
where 
\begin{equation}
\beta  \equiv {{\Omega _D } \mathord{\left/
 {\vphantom {{\Omega _D } {\Omega _H }}} \right.
 \kern-\nulldelimiterspace} {\Omega _H }} = \frac{{2\left( {1 + q} \right)}}{{a_ *  \left( {\chi ^3  + a_ *  } \right)}}
\end{equation}
Inspecting equations (15), (16), (20) and (21), we find the expressions for $P_{BZ}$ and $T_{BZ}$
are almost the same as those for $P_{MC}$ and $T_{MC}$ except that $k$ is replaced by $\beta$. From 
equation (9) we know that the parameter $k \equiv {{\Omega _F } \mathord{\left/
 {\vphantom {{\Omega _F } {\Omega _H }}} \right.
 \kern-\nulldelimiterspace} {\Omega _H }}
$ is uncertain due to lack of the 
knowledge about "the remote astrophysical load", while $\beta$ can be determined by equation (22) 
as a function of $a_*$ and $\chi$. 
}

\section{MAPPING RELATION BETWEEN BH HORIZON AND MC REGION OF DISC}
{
In order to calculate $P_{MC}$ and $T_{MC}$ we should first determine the mapping relation between 
the BH horizon and the disc. Considering the flux tube consisting of two adjacent magnetic 
surfaces "$i$ " and "$i+1$ " as shown in Fig.1, we have $\Delta \Psi_H=\Delta \Psi_D$ required by continuum of 
magnetic flux, i.e.,
\begin{equation}
B_H 2\pi ( \varpi \rho )_{r=r_H} d\theta  = - B_D 2\pi \left( {{{\varpi \rho } \mathord{\left/
 {\vphantom {{\varpi \rho } {\sqrt \Delta  }}} \right.
 \kern-\nulldelimiterspace} {\sqrt \Delta  }}} \right)_{\theta  = {\pi  \mathord{\left/
 {\vphantom {\pi  2}} \right.
 \kern-\nulldelimiterspace} 2}} dr
\end{equation}
where $B_D$ is the poloidal component of the magnetic field on the disc and
\begin{equation}
( \varpi \rho )_{r=r_H} = ( \Sigma \sin \theta )_{r=r_H} = 2Mr_H \sin \theta
\end{equation}
\begin{equation}
( \varpi \rho / \sqrt{\Delta} )_{\theta = \pi /2 } = \Sigma / \sqrt{\Delta} = \alpha ^{-1} \rho
\end{equation}
where
\begin{equation}
\alpha  = \left( {{{\rho \sqrt \Delta  } \mathord{\left/
 {\vphantom {{\rho \sqrt \Delta  } \Sigma }} \right.
 \kern-\nulldelimiterspace} \Sigma }} \right)_{\theta  = {\pi  \mathord{\left/
 {\vphantom {\pi  2}} \right.
 \kern-\nulldelimiterspace} 2}}  = \sqrt {\frac{{1 - 2\chi _{ms}^{ - 2} \xi ^{ - 1}  + a_ * ^2 \chi _{ms}^{ - 4} \xi ^{ - 2} }}{{1 + a_ * ^2 \chi _{ms}^{ - 4} \xi ^{ - 2}  + 2a_ * ^2 \chi _{ms}^{ - 6} \xi ^{ - 3} }}} 
\end{equation}
Following Blandford (1976) we assume that $B_D$ varies as 
\begin{equation}
B_D  \propto \xi ^{ - n} 
\end{equation}
where $\xi  \equiv r / {r_{ms} }$ is a dimensionless radial parameter defined in terms of the radius $r_{ms}$ 
of the marginally stable orbit (Bardeen et al 1972).
\\ \indent
Assuming that the inner boundary of the MC region is located at the inner edge of the disc, and 
considering the balance of the magneic pressures between the horizon and the inner edge of the 
disc (Ghosh \& Abramowicz 1997), we assume
\begin{equation}
B_H  = \left. {B_D } \right|_{\xi  = 1} 
\end{equation}
Incorporating equations (27) and (28) we have
\begin{equation}
B_D  = B_H \xi ^{ - n} 
\end{equation}
Substituting equations (24), (25) and (29) into equation (23) we have
\begin{equation}
\sin \theta d\theta  =  - {\rm{G}}\left( {a_ *  ;\xi ,n} \right)d\xi 
\end{equation}
where
\begin{equation}
{\rm{G}}\left( {a_ *  ;\xi ,n} \right) = \frac{{\chi _{ms}^4 \xi ^{ - n + 1} }}{{2(1 + q)}}\sqrt {\frac{{1 + a_ * ^2 \chi _{ms}^{ - 4} \xi ^{ - 2}  + 2a_ * ^2 \chi _{ms}^{ - 6} \xi ^{ - 3} }}{{1 - 2\chi _{ms}^{ - 2} \xi ^{ - 1}  + a_ * ^2 \chi _{ms}^{ - 4} \xi ^{ - 2} }}} 
\end{equation}
Integrating equation (30) we have
\begin{equation}
\cos \theta _M  - \cos \theta  =  - \int_{\xi _{out} }^\xi  {{\rm{G}}\left( {a_ *  ;\xi ,n} \right)} d\xi 
\end{equation}
Setting $\xi = \xi _{in} = 1$ at $\theta = \pi /2$, we have
\begin{equation}
\cos \theta _M  = \int_1^{\xi _{out} } {{\rm{G}}\left( {a_ *  ;\xi ,n} \right)} d\xi 
\end{equation}
Thus the mapping relation between the angular coordinate on the horizon and the radial coordinate on the disc 
is derived in the context of general relativity and conservation of magnetic flux as follows:
\begin{equation}
\cos \theta  = \int_1^\xi  {{\rm{G}}\left( {a_ *  ;\xi ,n} \right)} d\xi 
\end{equation}
It is noted that equation (33) provides a constraint to the outer boundary of the MC region, which is represented by 
$ \xi _{out} \equiv r_{out} / r_{ms}$, provided that the parameters $a_*$, $\theta _M$ and $n$ are given. $\xi _{out}$ 
turns out to be a function of $n$, varying monotonically for the given $a_*$ and $\theta _M$ as shown in Fig.3, and 
it behaves as a function of $a_*$ varying non-monotonically for the given $n$ and $\theta _M$ as shown in Fig.4. 
Since $\xi _{out}$ is limited, the parameter $n$ should be less than one critical value $n_c$ corresponding to infinite 
$\xi _{out}$. The curves of $n_c$ versus $a_*$ for the given values of $\theta _M$ are shown in Fig.5.
\begin{figure}
  {\includegraphics[width=6cm]{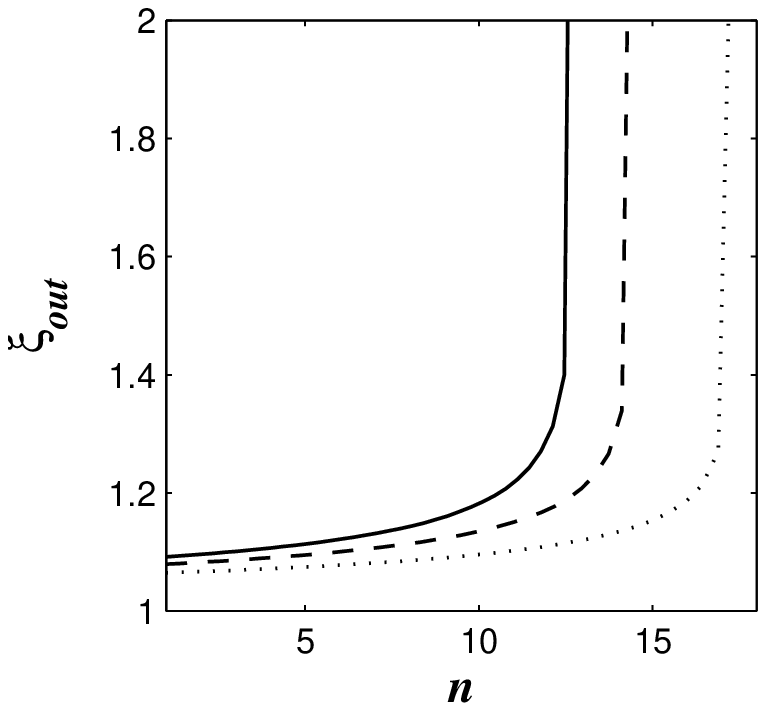}}
  \centerline{(a)\hspace{1.3cm}}\\ \\ \\
  {\includegraphics[width=6cm]{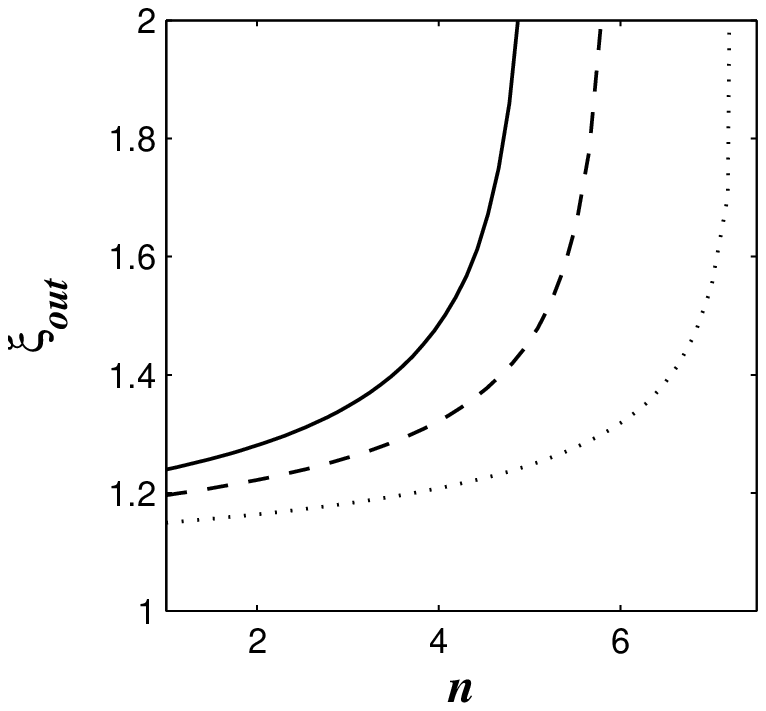}}
  \centerline{(b)\hspace{1.3cm}}\\
  \caption{The curves of $\xi _{out}$ versus $n$ for the given values of $\theta _M$ with $\theta _M =0$ (solid line), $\pi /6$ 
(dashed line) and $\pi /4$ (dotted line); (a) $a_* =0$ , (b) $a_* = 0.998$.}
\end{figure}
 \begin{figure}
  {\includegraphics[width=6cm]{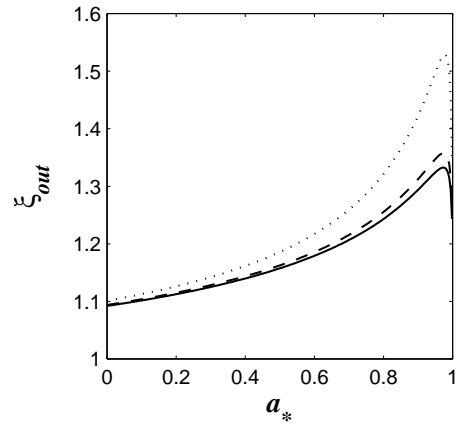}}
  \caption{The curves of $\xi _{out}$ versus $a_*$ for $\theta_M=0$ and $0<a_* <0.998$ with $n=1.1$ 
(solid line), $n=1.5$ (dashed line), and $n=3$ (dotted line).}
\end{figure}
\begin{figure}
  {\includegraphics[width=6cm]{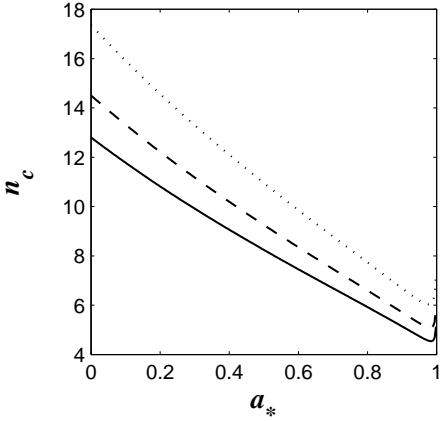}}
  \caption{The curves of $n_c$ versus $a_*$ for the given values of $\theta _M$ with $\theta _M =0$ (solid line), $\pi /6$ 
(dashed line) and $\pi /4$ (dotted line).}
\end{figure}
}

\section{COMPARISON OF POWER AND TORQUE IN THE BZ PROCESS AND MC PROCESS}
{
In order to compare the relative strength of MC power to the BZ power and torque we define the ratio 
of $P_{MC}$ to $P_{BZ}^{optimal}$ and that of $T_{MC}$ to $T_{BZ}^{optimal}$ as follows, 
\begin{eqnarray}
\gamma & \equiv & {{P_{MC} } \mathord{\left/
 {\vphantom {{P_{MC} } {P_{BZ}^{optimal} }}} \right.
 \kern-\nulldelimiterspace} {P_{BZ}^{optimal} }} \nonumber  \\ 
& = & \frac{{2a_ * ^2 Q}}{{\left( {arctanQ - {{a_ *  } \mathord{\left/
 {\vphantom {{a_ *  } 2}} \right.
 \kern-\nulldelimiterspace} 2}} \right)}}\int_{\theta _M }^{{\pi  \mathord{\left/
 {\vphantom {\pi  2}} \right.
 \kern-\nulldelimiterspace} 2}} {\frac{{\beta \left( {1 - \beta } \right)\sin ^3 \theta d\theta }}{{2 - \left( {1 - q} \right)\sin ^2 \theta }}} 
\end{eqnarray}
\begin{eqnarray}
\tau & \equiv & {{T_{MC} } \mathord{\left/
 {\vphantom {{T_{MC} } {T_{BZ}^{optimal} }}} \right.
 \kern-\nulldelimiterspace} {T_{BZ}^{optimal} }} \nonumber \\
& = & \frac{{a_ * ^2 Q}}{{\left( {arc\tan Q - {{a_ *  } \mathord{\left/
 {\vphantom {{a_ *  } 2}} \right.
 \kern-\nulldelimiterspace} 2}} \right)}}\int_{\theta _M }^{{\pi  \mathord{\left/
 {\vphantom {\pi  2}} \right.
 \kern-\nulldelimiterspace} 2}} {\frac{{\left( {1 - \beta } \right)\sin ^3 \theta d\theta }}{{2 - \left( {1 - q} \right)\sin ^2 \theta }}} 
\end{eqnarray}
Setting $\theta _M =0$ in equations (35) and (36), we have the curves of $\gamma$ and $\tau $ versus $a_* $ for the given 
parameter $n$, and those versus $n$ for the given $a_* $ as shown in Fig.6 and Fig.7, respectively.
 \begin{figure}
  {\includegraphics[width=6cm]{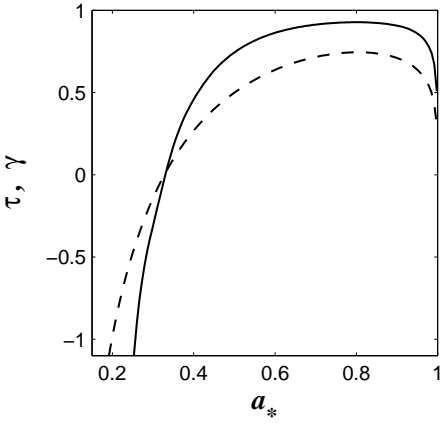}}
  \caption{The curves of $\gamma$ (solid line) and $\tau$ (dashed line) versus the BH spin $a_*$ for $n = 1.1$ and $0.2 < a_ *   < 0.998$.}
\end{figure}

\begin{table}
\caption{The three special values of $\gamma$ and the values of the concerning parameters}
\begin{tabular}{cccc}
\hline
$n$ & $\gamma =-1$ & $\gamma =0$ & $\gamma =\gamma _{max}$ \\ \hline
$1.1$ & $a_ * ^{\gamma 1}  = {\rm{0}}{\rm{.2583}}$ & $a_ * ^{\gamma 2}  = {\rm{0}}{\rm{.3286}}$ & 
$a_ * ^{\gamma m}  = 0.{\rm{7997}}$, \\
& & & $\gamma _{\max }  = 0.9278$ \\ 
$1.5$ & $a_ * ^{\gamma 1}  = {\rm{0}}{\rm{.2581}}$ & $a_ * ^{\gamma 2}  = {\rm{0}}{\rm{.3283}}$ &
$a_ * ^{\gamma m}  = 0.{\rm{8002}}$, \\
& & & $\gamma _{\max }  = 0.9287$ \\ 
$3.0$ & $a_ * ^{\gamma 1}  = {\rm{0}}{\rm{.2572}}$ & $a_ * ^{\gamma 2}  = {\rm{0}}{\rm{.3269}}$ &
  $a_ * ^{\gamma m}  = 0.{\rm{8021}}$, \\
& & & $\gamma _{\max }  = 0.9325$ \\ \hline
\end{tabular}
\end{table}
\begin{table}
\caption{The three special values of $\tau$ and the values of the concerning parameters}
\begin{tabular}{cccc}
\hline
$n$ & $\tau =-1$ & $\tau =0$ & $\tau =\tau _{max}$ \\ \hline
$1.1$ & $a_ * ^{\tau 1}  = {\rm{0}}{\rm{.1988}}$ & $a_ * ^{\tau 2}  = {\rm{0}}{\rm{.3278}}$ & 
  $a_ * ^{\tau m}  = 0.{\rm{8045}}$, \\ 
& & & $\tau _{\max }  = 0.7452$ \\ 
$1.5$ & $a_ * ^{\tau 1}  = {\rm{0}}{\rm{.1987}}$ & $a_ * ^{\tau 2}  = {\rm{0}}{\rm{.3275}}$ &
 $a_ * ^{\tau m}  = 0.{\rm{8056}}$, \\
& & & $\tau _{\max }  = 0.7481$ \\ 
$3.0$ & $a_ * ^{\tau 1}  = {\rm{0}}{\rm{.1980}}$ & $a_ * ^{\tau 2}  = {\rm{0}}{\rm{.3260}}$ &
  $a_ * ^{\tau m}  = 0.{\rm{8118}}$, \\
& & & $\tau _{\max }  = 0.7614$ \\ \hline
\end{tabular}
\end{table}

\begin{figure}
  \includegraphics[width=6cm]{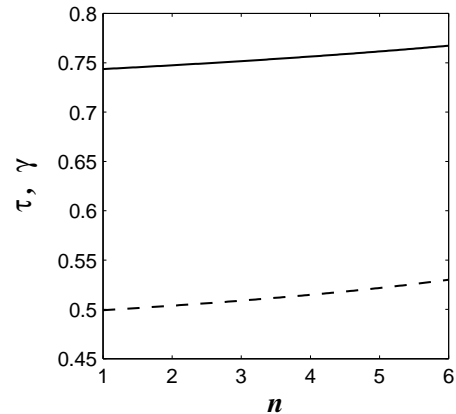}
  \caption{The curves of $\gamma$ (solid line) and $\tau$ (dashed line) versus the power-law index $n$ for $a_*  = 0.5$ and $1 < n < 6$.}
\end{figure}

The values of $a_*$ corresponding to the three special values of $\gamma$ ($\gamma=-1$, $0$ and $\gamma _{max}$) and 
to those of $\tau$($\tau=-1$, $0$ and $\tau _{max}$) are calculated for the given values of $n$ as shown in Table 1 and Table 2, 
respectively.

From the above calculations on $\gamma$ and $\tau$ we obtain the following results: \\ \indent
(i) The signs of $P_{MC} $ and $T_{MC} $ are the same as the corresponding $\gamma $ and $\tau$, and 
they change from the negative to the positive at $a_ * ^{\gamma 2}$ and $a_ * ^{\tau 2}$ respectively. 
It is shown in Table 1 and Table 2 that $a_ * ^{\tau 2}$ is a little less than $a_ * ^{\gamma 2}$. Therefore 
both energy and angular momentum are transferred from the BH to the disc as $a_ *   > a_ * ^{\gamma 2}$, 
and transferred from the disc to the BH as $a_ *   < a_ * ^{\tau 2}$. It is interesting to notice that the transfer 
direction of energy is opposite to that of angular momentum for $a_ * ^{\tau 2}  < a_ *   < a_ * ^{\gamma 2}$, 
i.e., energy is transferred from the disc to the BH, while angular momentum from the BH to the disc in this value 
range of the BH spin. \\ \indent
(ii) Both $\gamma$ and $\tau$ vary non-monotonically as $a_* $ for the given parameter $n$, and attain 
their maxima at $a_ * ^{\gamma m}$ and $a_ * ^{\tau m}$, respectively;  \\ \indent
(iii) The maxima of $P_{MC} $ and $T_{MC} $ are all less than the corresponding $P_{BZ}^{optimal}$ and 
$T_{BZ}^{optimal}$, respectively, while the absolute values of the former two are greater than the corresponding 
values of the latter two as $a_* $ is less than $a_ * ^{\gamma 1} $ and $a_ * ^{\tau 1} $, respectively. \\ \indent
(iv) As shown in Fig.7, both $\gamma$ and $\tau$ increase very little as the increasing parameter $n$, which implies 
that both $P_{MC} $ and $T_{MC} $ are insensitive to the power-law index $n$ of the radial coordinate $\xi$.
}

\section{EVOLUTION CHARACTERISTICS AND ENERGY EXTRACTING EFFICIENCY}
{
Based on conservation of energy and angular momentum the basic evolution equations of the 
Kerr BH in the coexistence of DABZMC are written as follows:
\begin{equation}
{{dM} \mathord{\left/
 {\vphantom {{dM} {dt}}} \right.
 \kern-\nulldelimiterspace} {dt}} = E_{ms} \dot M_D  - P_{BZ}  - P_{MC} 
\end{equation}
\begin{equation}
{{dJ} \mathord{\left/
 {\vphantom {{dJ} {dt}}} \right.
 \kern-\nulldelimiterspace} {dt}} = L_{ms} \dot M_D  - T_{BZ}  - T_{MC} 
\end{equation}
Incorporating equations (37) and (38), we have the evolution equation for the BH spin as 
follows:
\begin{eqnarray}
{{da_ *  } \mathord{\left/
 {\vphantom {{da_ *  } {dt}}} \right.
 \kern-\nulldelimiterspace} {dt}} & = & M^{ - 2} \left( {L_{ms} \dot M_D  - T_{BZ}  - T_{MC} } \right)  \nonumber \\ 
& &  - 2M^{ - 1} a_ *  \left( {E_{ms} \dot M_D  - P_{BZ}  - P_{MC} } \right)
\end{eqnarray}
where $\dot{M}_D$ is the accretion rate of rest mass, $E_{ms}$ and $L_{ms}$ are specific energy 
and specific  angular momentum corresponding to the inner edge radius $r_{ms}$, respectively. 

Since the magnetic field on the BH is supported by the surrounding disc, there are some 
relations between $B_H$ and $\dot{M}_D$. As a matter of fact these relations might be rather 
complicated, and would be very different in different situations. One of them is given to investigate the 
correlation between BH spin and dichotomy of quasars by considering the balance between the pressure 
of the magnetic field on the horizon and the ram pressure of the innermost parts of an accretion flow 
(Moderski, Sikora \& Lasota 1997), i.e., 
\begin{equation}
B_H^2 / (8 \pi ) = P_{ram} \sim \rho c^2 \sim \dot M_D / (4 \pi r_H^2 )
\end{equation}
From equation (40) we assume the relation as
\begin{equation}
\dot M_D  = {{B_H^2 M^2 \left( {1 + q} \right)^2 } \mathord{\left/
 {\vphantom {{B_H^2 M^2 \left( {1 + q} \right)^2 } 2}} \right.
 \kern-\nulldelimiterspace} 2}
\end{equation}
However equation (41) is derived without MC effects. By considering the transfer of the energy and angular 
momentum between the BH and the disc equations (41) is modified as follows:
\begin{equation}
\dot M_D  = {{B_H^2 M^2 \left( {1 + q} \right)^2 } \mathord{\left/
 {\vphantom {{B_H^2 M^2 \left( {1 + q} \right)^2 } 2}} \right.
 \kern-\nulldelimiterspace} 2} + \left[ {1 - \delta \left( {\theta _M  - {\pi  \mathord{\left/
 {\vphantom {\pi  2}} \right.
 \kern-\nulldelimiterspace} 2}} \right)} \right]\left( {\dot M_D } \right)_{mc} 
\end{equation}
where $\left( {\dot M_D } \right)_{mc}$ is the MC correction to the accretion rate at the inner edge of the disc, 
and it reads
\begin{eqnarray}
\lefteqn{\left( \dot M_D \right)_{mc}  = } \nonumber \\
& - \frac{{4B_H^2 M^2 Q\left( {\chi _{ms}^3  + a_ *  } \right)^2 \left( {1 - \beta } \right)}}{{\left( {\chi _{ms}^3  + 4a_ *  } \right)}}\sqrt {\frac{{1 + a_ * ^2 \chi _{ms}^{ - 4}  + 2a_ * ^2 \chi _{ms}^{ - 6} }}{{1 - 2\chi _{ms}^{ - 2}  + a_ * ^2 \chi _{ms}^{ - 4} }}} 
\end{eqnarray}
The derivation of equation (43) is given in Appendix A. To exclude $\left( {\dot M_D } \right)_{mc}$ from the 
case without magnetic coupling we define $\delta \left( {\theta _M  - {\pi  \mathord{\left/ {\vphantom {\pi  2}} \right.
 \kern-\nulldelimiterspace} 2}} \right) $ as follows:
\begin{equation}
\delta (\theta _M - \pi /2) \equiv \left\{ 
  \begin{array}{r@{\quad \quad}l}
  1, &  \theta _M = \pi /2, \\
  0, &  0 \le \theta _M < \pi /2
 \end{array}
\right.
\end{equation}
Thus the equation (42) is applicable to the coexistence of DABZMC. \\ \indent
Setting $\dot M_D  = 0 $ in equation (42), we can derive a value of the BH spin, $a_ * ^{stop}  \approx 0.4677$, 
at which the accretion will stop at the inner edge of the disc. Substituting equation (42) into equations (37)$-$(39), 
we have 
\begin{equation}
{{dM} \mathord{\left/
 {\vphantom {{dM} {dt}}} \right.
 \kern-\nulldelimiterspace} {dt}} = f \left( {a_ *  ,u_j } \right)\dot M_D 
\end{equation}
\begin{equation}
{{dJ} \mathord{\left/
 {\vphantom {{dJ} {dt}}} \right.
 \kern-\nulldelimiterspace} {dt}} = Mh \left( {a_ *  ,u_j } \right)\dot M_D 
\end{equation}
\begin{equation}
{{da_ *  } \mathord{\left/
 {\vphantom {{da_ *  } {dt}}} \right.
 \kern-\nulldelimiterspace} {dt}} = M^{ - 1} g \left( {a_ *  ,u_j } \right)\dot M_D 
\end{equation}
where 
\begin{equation}
g \left( {a_ *  ,u_j } \right) = h \left( {a_ *  ,u_j } \right) - 2a_ *  f \left( {a_ *  ,u_j } \right)
\end{equation}
In the above equations $u_j  \equiv \theta _M ,k, n$ is used for simplicity to represent the three 
parameters other than $a_*$. $f(a_*,u_j)$ and $g(a_*,u_j)$ are referred to as the characteristics functions 
(CFs) of BH evolution. 

\subsection{Evolution characteristics of Kerr BH}
First we discuss the evolution characteristics of the Kerr BH in the coexistence of disc accretion with the 
BZ process for $\theta_M= \pi / 2$. From equations (37)$-$(39) we have the CFs as follows:
\begin{equation}
f \left( {a_ *  ,k} \right) = E_{ms}  - {{P_{BZ} } \mathord{\left/
 {\vphantom {{P_{BZ} } {\dot M_D }}} \right.
 \kern-\nulldelimiterspace} {\dot M_D }}
\end{equation}
\begin{equation}
h \left( {a_ *  ,k} \right) = M^{ - 1} \left( {L_{ms}  - {{T_{BZ} } \mathord{\left/
 {\vphantom {{T_{BZ} } {\dot M_D }}} \right.
 \kern-\nulldelimiterspace} {\dot M_D }}} \right)
\end{equation}
\begin{equation}
g \left( {a_ *  ,k} \right) = h \left( {a_ *  ,k} \right) - 2a_ *  f \left( {a_ *  ,k} \right)
\end{equation}
From equations (45) and (47) we find that the signs of $dM /dt $ and $da_* /dt$ are the same as those of 
$f\left( {a_ *  ,k} \right)$ and $g\left( {a_ *  ,k} \right)$, respectively. As shown in Fig.8 we have the curve 
represented by $g\left( {a_ *  ,k} \right) = 0$ in the 2-dimension space consisting of the parameters $a_*$ and $k$, 
which divides the parameter space into two regions. And each black dot with arrowhead is referred to as a 
representative point (RP), which represents one BH evolution state. We can use RP's displacement in the parameter 
space to describe BH evolution visually. According to the positions of RPs we have two evolution modes of the BH, 
and the details are given as follows. \\ 
\\{\bf(I)  Region I:} $f(a_*,k)>0$ and $g(a_*,k)<0$. 
\\ \indent
{\bf Mode I:} RPs in Region I always move towards the left until they arrive at the boundary curve 
$g\left( {a_ *  ,k} \right) = 0$, i.e., BH will reach its equilibrium spin $a_*^{eq}$ in long enough evolution time. 
RPs in region I represent the BHs with increasing mass and decreasing spin.
\\{\bf(II)  Region II:} $f(a_*,k)>0$ and $g(a_*,k)>0$. 
\\ \indent
{\bf Mode II:} RPs in Region II always move towards the right until they arrive at the boundary curve 
$g\left( {a_ *  ,k} \right) = 0$, i.e., BH will reach its equilibrium spin $a_*^{eq}$ in long enough evolution time. 
RPs in region II represent the BHs with increasing mass and increasing spin.
\\ \indent
Next we shall discuss BH evolution in the coexistence of disc accretion with MC process. Substituting $\theta_M=0$
and equation (42) into equations (37)$-$(39), we have the corresponding CFs as follows:
\begin{equation}
f \left( {a_ *  ,n } \right) = E_{ms}  - {{P_{MC} } \mathord{\left/
 {\vphantom {{P_{MC} } {\dot M_D }}} \right.
 \kern-\nulldelimiterspace} {\dot M_D }}
\end{equation}
\begin{equation}
h \left( {a_ *  ,n } \right) = M^{ - 1} \left( {L_{ms}  - {{T_{MC} } \mathord{\left/
 {\vphantom {{T_{MC} } {\dot M_D }}} \right.
 \kern-\nulldelimiterspace} {\dot M_D }}} \right)
\end{equation}
\begin{equation}
g \left( {a_ *  ,n } \right) = h \left( {a_ *  ,n } \right) - 2a_ *  f \left( {a_ *  ,n } \right)
\end{equation}
 \begin{figure}
  {\includegraphics[width=6cm]{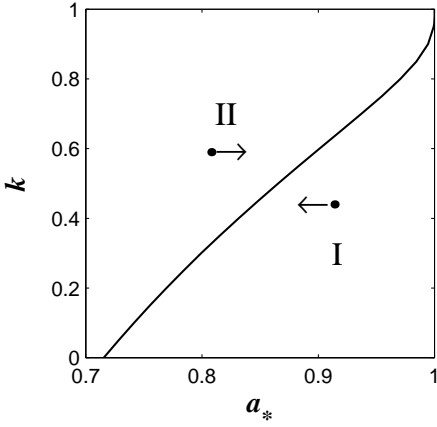}}
  \caption{Parameter space for BH evolution in the BZ process with $0<k<1$ and $0.7<a_* <1$.}
\end{figure}
\begin{figure}
  {\includegraphics[width=6cm]{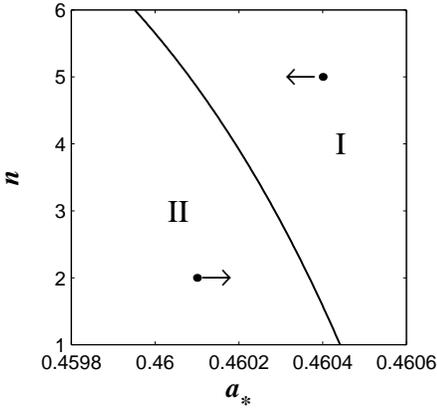}}
  \caption{Parameter space for BH evolution in MC process with $1<n<6$ and $0.4598 < a_ *   < 0.4606$.}
\end{figure}
We can also describe BH evolution by using the displacement of RPs in 2-dimension parameter space 
consisting of the parameters $a_*$ and $n$ as shown in Fig.9, which is divided into two regions by the 
curve $g\left( {a_ *  ,n} \right) = 0$. In this case the description of the BH evolution characteristics and 
the two evolution modes are almost the same as that given in the BZ process except that $f\left( {a_ *  ,k} \right) $ 
and $g\left( {a_ *  ,k} \right) $ are replaced by $f\left( {a_ *  ,n} \right) $ and $g\left( {a_ *  ,n} \right) $, 
respectively. Inspecting Fig.8 and Fig.9, we have the following results: \\ \indent
(i) The CFs in the corresponding parameter space can be used to describe the evolution characteristics of the BH 
in a visual way. If we know the initial position of RP in the parameter space, we can determine the evolution 
characteristics of the BH immediately. \\ \indent
(ii) The effects of the parameter $k$ and $n$ on the BH evolution in the BZ and MC process can be easily found 
in the corresponding parameter space. From Fig.8 we find that $a_*^{eq}$ increases as the increasing $k$, which 
is consistent with our previous work (WLY). On the other hand, we find from Fig.9 that $a_*^{eq}$ increases 
very little as the decreasing $n$. Our calculations show that the values of $a_*^{eq}$ are very near to and a little 
less than $a_ * ^{stop}  \approx 0.4677$. \\ \indent
(iii) As shown in Fig.5, the value range of $n$ is limited and confined by conservation of magnetic flux. Compared 
with the parameter $k$, the effects of $n$ on the variation of $a_*^{eq}$ are very weak for the confined value 
range of $n$.

\subsection{Energy extracting efficiency of BH accretion disc}
Energy extracting efficiency of BH accretion disc can be investigated also by using the CFs. From equations (45) 
and (47) we have
\begin{equation}
{{dM} \mathord{\left/
 {\vphantom {{dM} M}} \right.
 \kern-\nulldelimiterspace} M} = \left[ {{{f \left( {a_ *  ,u_j } \right)} \mathord{\left/
 {\vphantom {{f \left( {a_ *  ,u_j } \right)} {g \left( {a_ *  ,u_j } \right)}}} \right.
 \kern-\nulldelimiterspace} {g \left( {a_ *  ,u_j } \right)}}} \right]da_ *  
\end{equation}
As is well known, a Kerr BH can be characterized by its mass $M$ and spin $a_*$. If a Kerr BH 
evolves from the initial state $(a_{*0},M_0)$ to the state $(a_*,M)$, the BH mass can be expressed 
by
\begin{equation}
M = M_0 \exp \left[ {\int_{a_{ * 0}}^{a_*} \frac{f(a_{*}^{'},u_j)}{g(a_{*}^{'},u_j)}} da_{*}^{'} \right].
\end{equation}
The efficiency corresponding to the evolution process of BH can be defined as
\begin{equation}
\eta _p  = \frac{{M_D  - \left( {M - M_0 } \right)}}{{M_D }} = 1 - \frac{{{M \mathord{\left/
 {\vphantom {M {M_0 }}} \right.
 \kern-\nulldelimiterspace} {M_0 }} - 1}}{{{{M_D } \mathord{\left/
 {\vphantom {{M_D } {M_0 }}} \right.
 \kern-\nulldelimiterspace} {M_0 }}}}
\end{equation}
which is the efficiency of converting accreted mass into radiation energy during the evolution process, and $M_D$ 
is the accreted rest mass and expressed as a function of $a_*$ by using equations (47) and (56) as follows:
\begin{equation}
M_D  = M_0 \int_{a_{ * 0} }^{a_ *  } {\frac{1}{{g \left( {a_ *  ,u_j } \right)}}} \exp \left[ {\int_{a_{ * 0} }^{a_ *  } 
{\frac{{f \left( {a_ {*} ^{'} ,u_j } \right)}}{{g \left( {a_ {*} ^{'} ,u_j } \right)}}da_ {*} ^{'} } } \right]da_ *  
\end{equation}
The efficiency corresponding to each evolving state of the BH is defined as
\begin{equation}
\eta _s  = \frac{{dM_D  - dM}}{{dM_D }} = 1 - {{\left( {{{dM} \mathord{\left/
 {\vphantom {{dM} {dt}}} \right.
 \kern-\nulldelimiterspace} {dt}}} \right)} \mathord{\left/
 {\vphantom {{\left( {{{dM} \mathord{\left/
 {\vphantom {{dM} {dt}}} \right.
 \kern-\nulldelimiterspace} {dt}}} \right)} {\dot M_D }}} \right.
 \kern-\nulldelimiterspace} {\dot M_D }} = 1 - f \left( {a_ *  ,u_j } \right)
\end{equation}
From equation (37) we find that efficiency $\eta_s$ is equal to the sum of the following three terms:
\begin{equation}
\eta _s  = \eta _{DA}  + \eta _{BZ}  + \eta _{MC} 
\end{equation}
where  $\eta_{DA}=1-E_{ms}$, $\eta_{BZ}=P_{BZ}/ \dot{M}_D$ and $\eta_{MC}=P_{MC}/ \dot{M}_D$ 
are the contributions of disc accretion, the BZ process and MC process, respectively. 
From equation (47) we know that the equilibrium spin $a_{*}^{eq}$ is determined by
\begin{equation}
g \left( {a_ *  ,u_j } \right) = 0
\end{equation}
Incorporating equations (55)$-$(59), we have
\begin{equation}
\left. {\eta _p } \right|_{a_ *   \to a_ * ^{eq} }  = \mathop {\lim }\limits_{a_ *   \to a_ * ^{eq} } \left( {1 - {A \mathord{\left/
 {\vphantom {A B}} \right.
 \kern-\nulldelimiterspace} B}} \right)
\end{equation}
In equation (62) we have
$$A  \equiv \exp \int_{a_{ * 0} }^{a_ *  } {\frac{{f \left( {a_ {*} ^{'} ,u_j } \right)}}{{g \left( {a_ {*} ^{'} ,u_j } \right)}}da_{ *} ^{'} }  - 1 \to \infty $$
and
$$B  \equiv \int_{a_{ * 0} }^{a_ *  } {\frac{1}{{g \left( {a_ *  ,u_j } \right)}}} 
\exp \left[ {\int_{a_{ * 0} }^{a_ *  } {\frac{{f \left( {a_{ *} ^{'} ,u_j } \right)}}{{g \left( {a_ {*} ^{'} ,u_j } \right)}}da_ {*} ^{'} } } \right]da_ *   \to \infty $$
By using L' Hospital limit rule for $\infty/ \infty$ type we have
\begin{equation}
\left. {\eta _p } \right|_{a_ *   \to a_ * ^{eq} }  = 1 - f \left( {a_ * ^{eq} ,u_j } \right) = \eta _s^{eq} \left( {a_ * ^{eq} ,u_j } \right)
\end{equation}
where $\eta_{s}^{eq}$ represents the efficiency $\eta_s$ at $a_{*}^{eq}$. Equation (63) implies that the efficiency $\eta_p$
is exactly equal to $\eta_{s}^{eq}$, provided that a BH evolves from any spin $a_{*0}$ to its equilibrium spin $a_{*}^{eq}$. 
Incorporating equations (42), (59), (61) and (37)$-$(39), we obtain the curves of $a_{*}^{eq}$ and $\eta_{s}^{eq}$ varying 
as the concerning parameters are shown in Fig.10.
\begin{figure}{
 \resizebox{\hsize}{!}{
{\includegraphics[width=6cm]{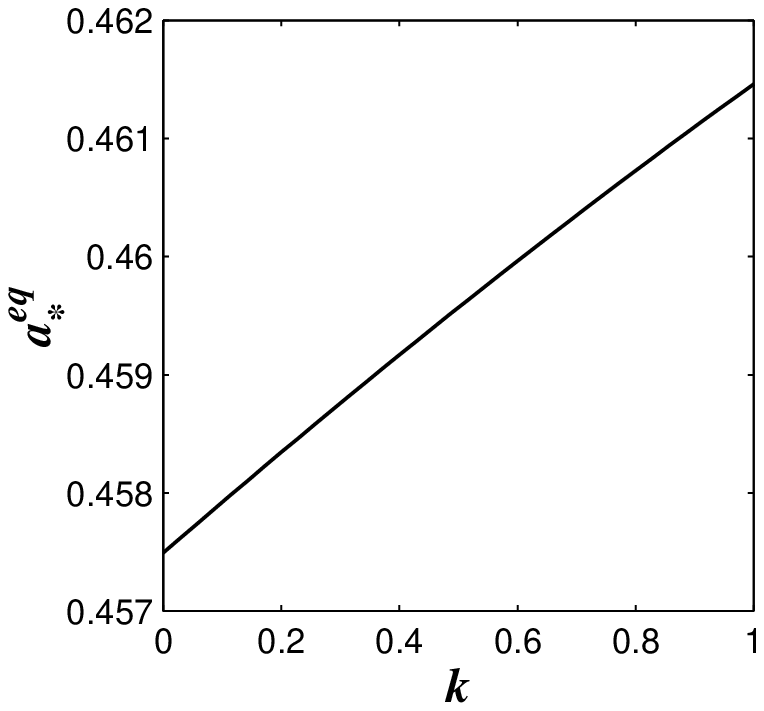}}(a)\ \ 
{\includegraphics[width=6cm]{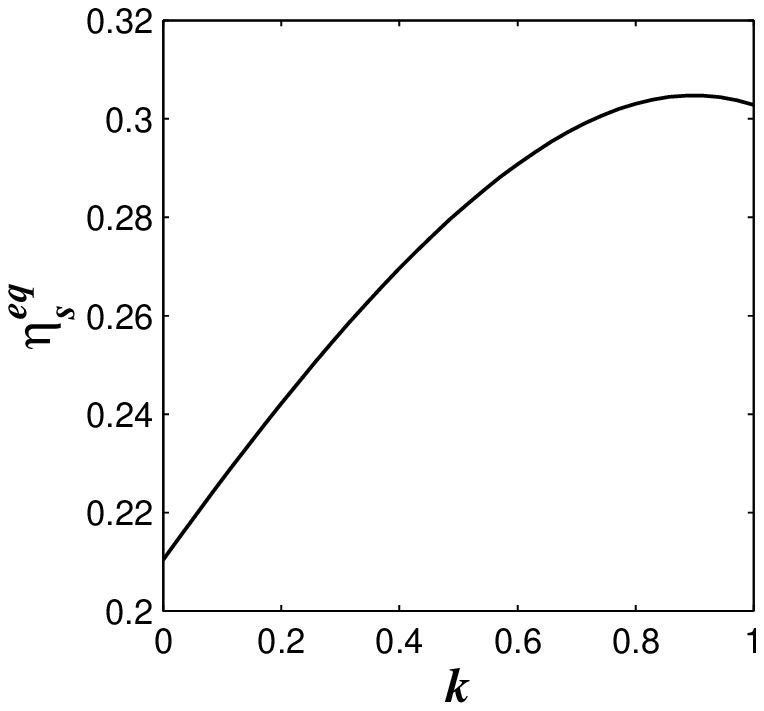}}(d)
}\\
 \resizebox{\hsize}{!}{
{\includegraphics[width=6cm]{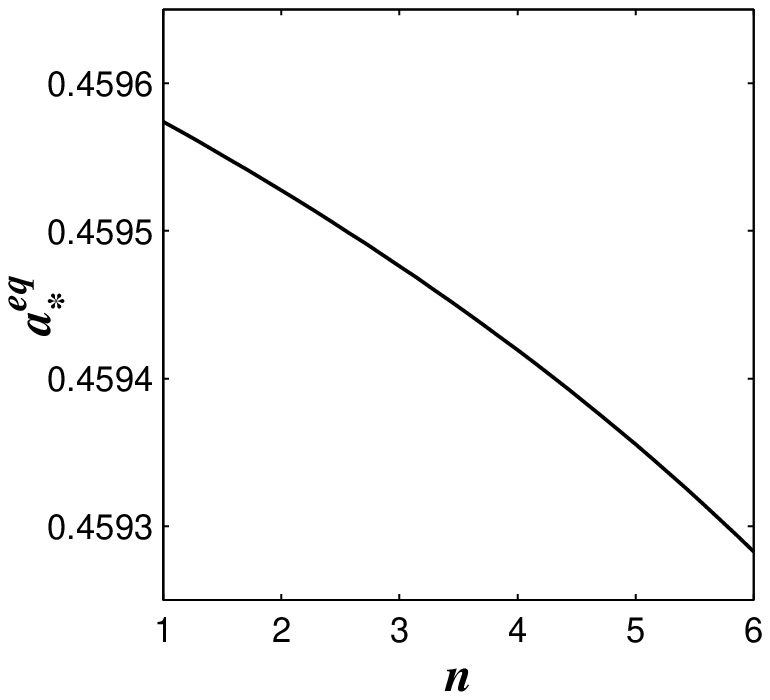}}(b)\ \ 
{\includegraphics[width=6cm]{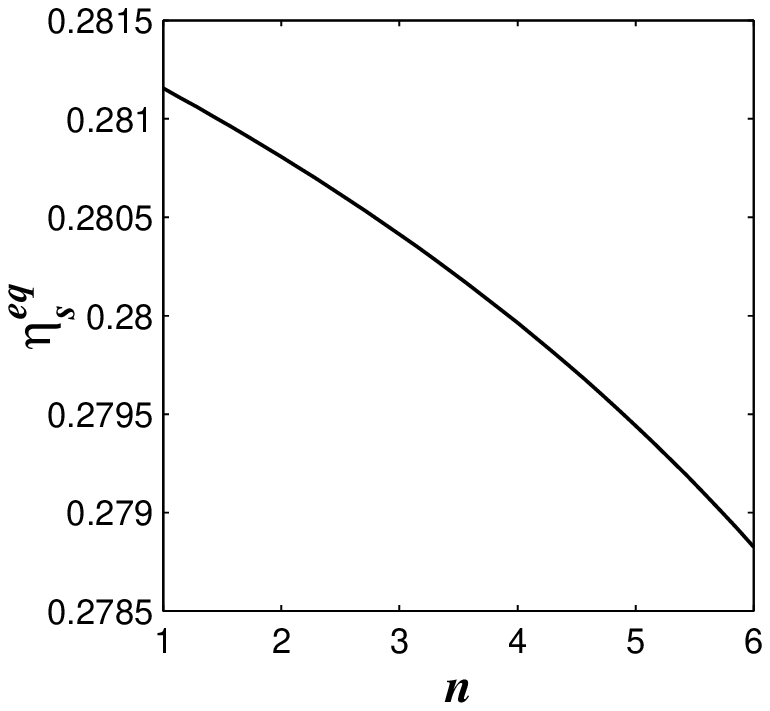}}(e)
}\\ \ \ \\
 \resizebox{\hsize}{!}{
{\includegraphics[width=6cm]{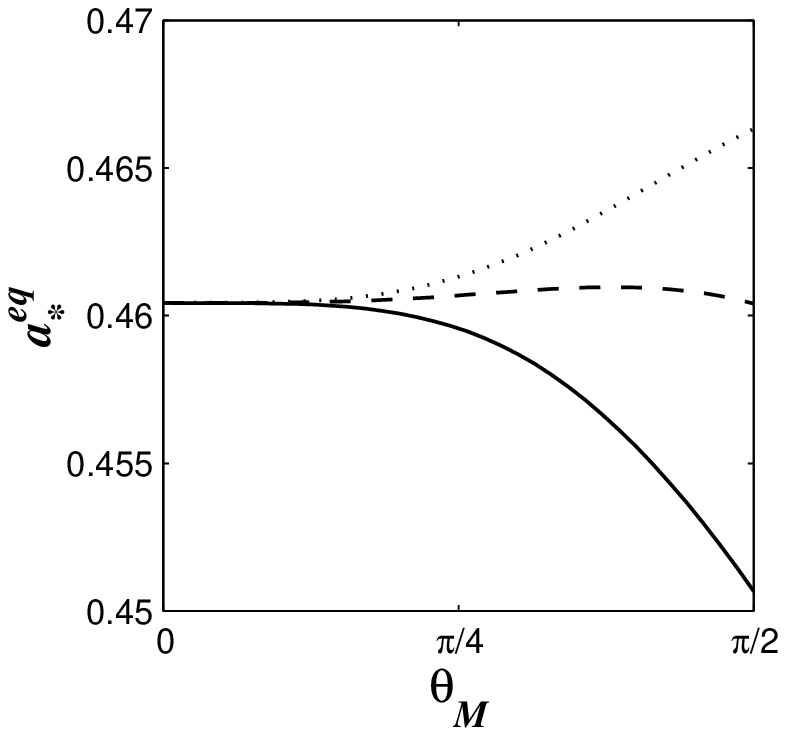}}(c)\ \ 
{\includegraphics[width=6cm]{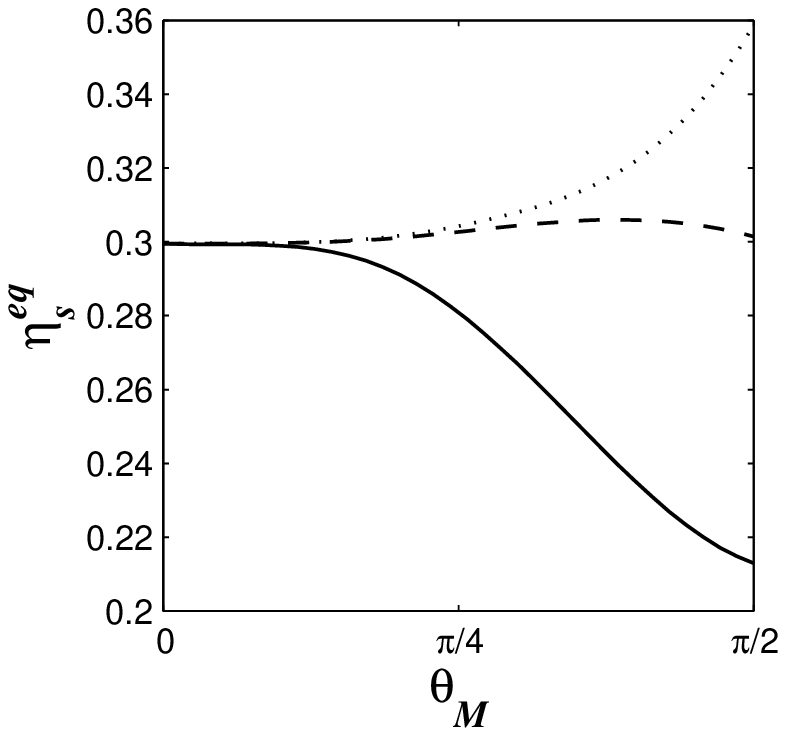}}(f)
}
}
\caption{The curves of $a_{*}^{eq}$ and $\eta_{s}^{eq}$ varying as the concerning parameters.
(a), (d) $\theta_M=\pi /4$, $0<k<1$, $n=1.1$; (b), (e) $\theta_M=\pi /4$, $k=0.5$, $1<n<6$; (c), (f) 
$0<\theta_M<\pi /2$, $n=1.1$, $k=0.5$(solid line), $k=0.788$(dashed line); $k=0.96$(dotted line).}
  \label{fig:fig8}
\end{figure}

Comparing Fig.10a, b, c with Fig.10d, e, f we find the following results:

(i) The curves of $a_{*}^{eq}$ and $\eta_{s}^{eq}$ varying as the concerning parameters look very alike 
except that $\eta_{s}^{eq}$ varies as $k$ non-monotonically with its maximum $(\eta_{s}^{eq})_{max} \approx 30.5 \%$ at  
$a_{*}^{eq} \approx 0.4611$ for $k \approx 0.8998$ as shown in Fig.10d. 

(ii) Both $a_{*}^{eq}$ and $\eta_{s}^{eq}$ decrease monotonically as the increasing $n$ for $\theta_M=\pi /4$ 
as shown in Fig.10b and Fig.10e. Considering that the outer boundary $\xi _{out}$ increase monotonically as the increasing 
$n$, we infer that the more concentrated is the magnetic field on the central region of the disc, the wider is the MC region 
on the disc, the stronger is the braking effects of MC process on the BH spin, and the lower are the $a_{*}^{eq}$ and 
$\eta_{s}^{eq}$.

(iii) As shown in Figs.10c and 10f the variations of $a_{*}^{eq}$ and $\eta_{s}^{eq}$ as $\theta_M$ display different 
characteristics for different values of parameter $k$: They might increase as $\theta_M$ 
(dotted line) or decreases as $\theta_M$ (solid line) or almost remain unchanged with $\theta_M$ 
(dashed line). It is not difficult to understand these results by considering that $\theta_M$ is the angular 
boundary between the open and closed field lines on the BH horizon: The different variations of $a_{*}^{eq}$ 
and $\eta_{s}^{eq}$ as $\theta_M$ come from the different contributions due to the BZ and MC process. 
For example, $a_{*}^{eq}$ and $\eta_{s}^{eq}$ will increase as $\theta_M$ , if the contributions of the open field lines in 
unit angular coordinate are greater than those corresponding to the closed ones, while they will 
decrease as $\theta_M$ in case that the former is less than the latter. 

In summary we can calculate the efficiency in an evolution process and that corresponding to each evolving state 
of the Kerr BH by using the equations (57) and (59), respectively, and both of which are expressed in terms of the CFs.  
}

\section{BH ENTROPY CHANGE IN COEXISTENCE OF DABZMC}
{
It was shown that the excess rate of change of entropy of the central BH in the BZ process comes from the total 
power dissipation on the BH horizon (WLY). It is attractive for us to extend this result to the coexistence of DABZMC. 
As is well known, the BH temperature $T_H$ and entropy $S_H$ can be expressed as 
(Thorne, Price \& Macdonald 1986, hereafter TPM)
\begin{equation}
T_H  = \frac{q}{{4\pi M(1 + q)}},
S_H  = 2\pi M^2 (1 + q)
\end{equation}
Incorporating the basic evolution equations (37), (38) and (39), we have the following equations,
\begin{equation}
\begin{array}{l}
 T_H { {dS_H }/{dt}} = {{dM}/{dt}} - \Omega _H { {dJ}/ {dt}} =  
 ( E_{ms}  -  \\ 
 \Omega _H L_{ms} ) \dot M_D 
+ \left( {\Omega _H T_{BZ}  - P_{BZ} } \right) + \left( {\Omega _H T_{MC}  - P_{MC} } \right) \\ 
 \end{array}
\end{equation}
equation (65) is exactly the mathematical formulation of the first law of thermodynamics for a Kerr BH (TPM). 
Obviously, the first term on RHS of equation (65) is the contribution due to disc accretion. It is easy to prove 
that the second and the third term arise from power dissipation on the BH horizon due to the BZ and MC process, 
respectively. By using the improved equivalent circuit as shown in Fig.2 and equation (11) we have 
\begin{equation}
I\Delta \varepsilon _H  = I^2 \Delta Z_H  + 
I^2 \Delta Z_L  - I\Delta \varepsilon _L 
\end{equation}
Incorporating equations (9), (11), (13) and 
$\Delta \varepsilon _L  =  - \left( {{{\Delta \Psi } \mathord{\left/
 {\vphantom {{\Delta \Psi } {2\pi }}} \right.
 \kern-\nulldelimiterspace} {2\pi }}} \right)\Omega _L $ we obtain 
\begin{equation}
\Delta P = \left( {{{\Delta \Psi } \mathord{\left/
 {\vphantom {{\Delta \Psi } {2\pi }}} \right.
 \kern-\nulldelimiterspace} {2\pi }}} \right)^2 
\frac{{\Omega _F \left( {\Omega _H  - \Omega _F } \right)}}
{{\Delta Z_H }} = I^2 \Delta Z_L  - I\Delta \varepsilon _L 
\end{equation}
From equations (66) and (67) we find that $I \Delta \varepsilon_H$ consists of two parts: the extracting 
power $\Delta P$ and the dissipated power $\Delta P_{DSP}$. 
The extracting power is transferred to the load including both the 
remote astrophysical load and the load disc.
For the BZ process  $\Delta \varepsilon_L =0$, we have
\begin{equation}
\Delta P = \Delta P_{BZ}  = I^2 \Delta Z_L  = 
\left( {{{\Delta \Psi } \mathord{\left/
 {\vphantom {{\Delta \Psi } {2\pi }}} \right.
 \kern-\nulldelimiterspace} {2\pi }}} \right)^2 
\frac{{\Omega _F \left( {\Omega _H  - \Omega _F } \right)}}{{\Delta Z_H }}
\end{equation}
For MC process $\Delta Z_L=0$, we have 
\begin{equation}
\Delta P = \Delta P_{MC}  =  - I\Delta \varepsilon _L  = 
\left( {{{\Delta \Psi } \mathord{\left/
 {\vphantom {{\Delta \Psi } {2\pi }}} \right.
 \kern-\nulldelimiterspace} {2\pi }}} \right)^2 
\frac{{\Omega _D \left( {\Omega _H  - \Omega _D } \right)}}{{\Delta Z_H }}
\end{equation}
Incorporating equations (11), (12) and (13), we can express the 
dissipated power $\Delta P_{DSP}$ on the BH horizon as
\begin{eqnarray}
\lefteqn{ \Delta P_{DSP}  = I^2 \Delta Z_H  = {{\left( {{{\Delta \Psi } \mathord{\left/
 {\vphantom {{\Delta \Psi } {2\pi }}} \right.
 \kern-\nulldelimiterspace} {2\pi }}} \right)^2 \left( {\Omega _H  - \Omega _F } \right)^2 } \mathord{\left/
 {\vphantom {{\left( {{{\Delta \Psi } \mathord{\left/
 {\vphantom {{\Delta \Psi } {2\pi }}} \right.
 \kern-\nulldelimiterspace} {2\pi }}} \right)^2 \left( {\Omega _H  - \Omega _F } \right)^2 } {\Delta Z_H }}} \right.
 \kern-\nulldelimiterspace} {\Delta Z_H }} } \nonumber \\ 
\lefteqn{= \Omega _H \Delta T - \Delta P  }
\end{eqnarray}
Integrating equation (70) over the angular coordinate $\theta$, 
we obtain the total dissipated power on the horizon, 
which is exactly equal to the sum of the second and the third term on RHS of 
equation (65), i.e.,
\begin{equation}
P_{DSP}  = \left( {P_{DSP} } \right)_{BZ}  + \left( {P_{DSP} } \right)_{MC} 
\end{equation}
where $\left( {P_{DSP} } \right)_{BZ}  = 
\left( {\Omega _H T_{BZ}  - P_{BZ} } \right)$ and 
$\left( {P_{DSP} } \right)_{MC}  = 
\left( {\Omega _H T_{MC}  - P_{MC} } \right)$ are the total power 
dissipation due to the BZ and MC process, respectively. 
Combining equations (15), (16), (20) and (21) we have
\begin{eqnarray}
\left( {P_{DSP} } \right)_{BZ} & = & \Omega _H T_{BZ}  - P_{BZ}  \nonumber  \\ 
 & = & 2a_ * ^2 B_H^2 M^2 \int_0^{\theta _M } 
{\frac{{\left( {1 - k} \right)^2 \sin ^3 \theta d\theta }}
{{2 - \left( {1 - q} \right)\sin ^2 \theta }}} 
\end{eqnarray}
\begin{eqnarray}
\left( {P_{DSP} } \right)_{MC} & = & \Omega _H T_{MC}  - P_{MC}  \nonumber \\
 & = & 2a_ * ^2 B_H^2 M^2 \int_{\theta _M }^{{\pi  \mathord{\left/
 {\vphantom {\pi  2}} \right.
 \kern-\nulldelimiterspace} 2}} 
{\frac{{\left( {1 - \beta } \right)^2 \sin ^3 \theta d\theta }}
{{2 - \left( {1 - q} \right)\sin ^2 \theta }}}  
\end{eqnarray}
Incorporating equations (65), (72) and (73), we express the rates of change of BH entropy corresponding to disc 
accretion, the BZ and MC process as follows:
\begin{equation}
\left( {{{dS_H } \mathord{\left/
 {\vphantom {{dS_H } {dt}}} \right.
 \kern-\nulldelimiterspace} {dt}}} \right)_{DA} 
 = T_H^{ - 1} \left( {E_{ms}  - \Omega _H L_{ms} } \right)\dot M_D 
\end{equation}
\begin{equation}
\left( {{{dS_H } \mathord{\left/
 {\vphantom {{dS_H } {dt}}} \right.
 \kern-\nulldelimiterspace} {dt}}} \right)_{BZ}  = 
T_H^{ - 1} \left( {\Omega _H T_{BZ}  - P_{BZ} } \right)
\end{equation}
\begin{equation}
\left( {{{dS_H } \mathord{\left/
 {\vphantom {{dS_H } {dt}}} \right.
 \kern-\nulldelimiterspace} {dt}}} \right)_{MC}  = 
T_H^{ - 1} \left( {\Omega _H T_{MC}  - P_{MC} } \right)
\end{equation}
From the above analysis we conclude that the excess rate of change of the BH entropy does come 
from the total power dissipation on the BH horizon in the BZ and MC process.

The ratios of the rates of BH entropy change can be written by
\begin{eqnarray}
\lefteqn{  R_{BZDA} \left( {a_ *  ,\theta _M ,k} \right) =\frac {({{dS_H }/{dt}})_{BZ}}
    {({{dS_H }/{dt}})_{DA}} } \nonumber \\
\lefteqn{= \frac{{\Omega _H T_{BZ}  - P_{BZ} }}{{\left( {E_{ms}  - \Omega _H L_{ms} } \right)\dot M_D }} }
\end{eqnarray}
\begin{eqnarray}
\lefteqn{ R_{MCDA} \left( {a_ *  ,\theta _M , n} \right) = \frac{{\left( {{{dS_H } \mathord{\left/
 {\vphantom {{dS_H } {dt}}} \right.
 \kern-\nulldelimiterspace} {dt}}} \right)_{MC} }}{{\left( {{{dS_H } \mathord{\left/
 {\vphantom {{dS_H } {dt}}} \right.
 \kern-\nulldelimiterspace} {dt}}} \right)_{DA} }} } \nonumber \\ 
\lefteqn{= \frac{{\Omega _H T_{MC}  - P_{MC} }}{{\left( {E_{ms}  - \Omega _H L_{ms} } \right)\dot M_D }} }
\end{eqnarray}
where the four parameters $k$, $a_*$, $\theta_M$ and $n$ are involved. Letting one of them vary 
and the rest be fixed, we have the curves of $R_{BZDA}$ and $R_{MCDA}$ 
versus $a_*$ and $\theta _M$ as shown in Fig.11, and the following results are obtained:

(i) As shown in Figs.11a and 11b $R_{BZDA}$ increases monotonically as 
$a_*$ for the given values of $k$ and $\theta _M$, while $R_{MCDA}$ varies non-monotonically and attains 
a minimum approach zero at $a_*  \approx {\rm{0}}{\rm{.3305}}$ for the given values of $n$ and $\theta _M$. 
From the discussion in Sec.IV we know that the minimum of $R_{MCDA}$ arises from the neighboring $a_ * ^{\gamma 2}$ 
and $a_ * ^{\tau 2}$ corresponding to $P_{MC}  = 0$ and $T_{MC}  = 0$, respectively.

(ii) As shown in Figs.11a and 11b we have $R_{MCDA}>R_{BZDA}$ for $0<a_*<0.2694$ and $0.4373<a_*<a_*^{eq}$, 
which implies that the contribution to $P_{DSP}$ and BH entropy increase 
due to MC process dominates over that due to the BZ process in the above value range of the BH spin.

(iii) As shown in Fig.11c the ratio $R_{BZDA}$ increases monotonically as $\theta _M$ for the given values of $a_*$ and $k$, 
while $R_{MCDA}$ varies non-monotonically as $\theta _M$ for the above given values of $a_*$ and $n$. $R_{MCDA}$ is 
greater than $R_{BZDA}$ for $0 < \theta _M  < 0.{\rm{6782}}$, while the former is much less than the latter for $0.6782 < \theta _M  < {\pi  \mathord{\left/
 {\vphantom {\pi  2}} \right.
 \kern-\nulldelimiterspace} 2}$. It is obvious that the sum of the two is much less than unity, which implies that the contribution 
to BH entropy increase is dominated by disc accretion in the coexistence of DABZMC. 
\begin{figure}
  {\includegraphics[width=5.6cm]{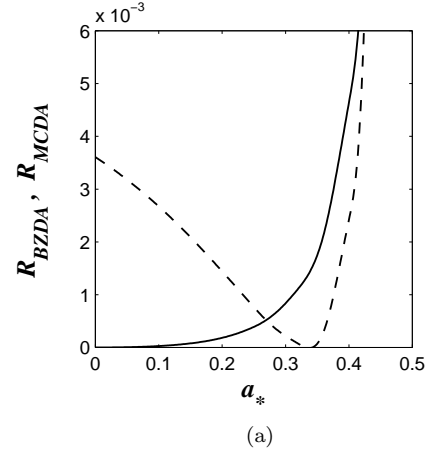}}
  \centerline{(a)\hspace{1.8cm}}\\ \\
  {\includegraphics[width=5.6cm]{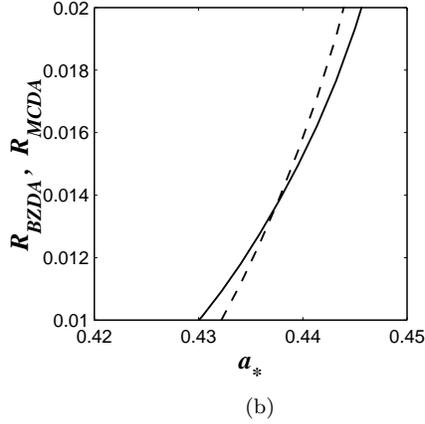}}
  \centerline{(b)\hspace{1.8cm}}\\ \\
  {\includegraphics[width=5.6cm]{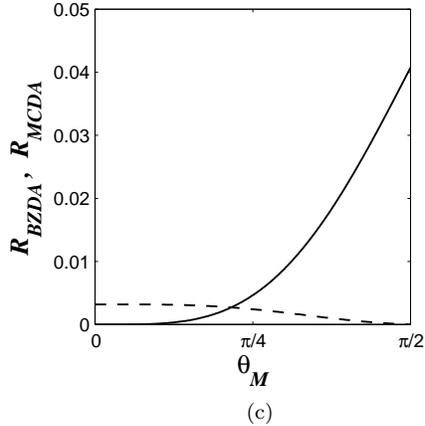}}
  \centerline{(c)\hspace{1.8cm}}\\
  \caption{The curves of $R_{BZDA}$ (solid line) and $R_{MCDA}$(dashed line) versus the concerning parameters,
(a), (b) $0<a_*<1$, $\theta_M=\pi/4$, $k=0.5$, $n=1.1$;
(c) $0<\theta_M<\pi/2$, $a_*=0.4$, $k=0.5$, $n=1.1$.}
\end{figure}

}

\section{SUMMARY}
{
In this paper the evolution characteristics of a Kerr BH and some related issues, such as energy extracting efficiency 
and entropy change on the BH horizon, are investigated by considering coexistence of DABZMC. By using an improved 
equivalent circuit a unified expression for the BZ power and MC power are derived. Starting from the conservation 
laws of energy and angular momentum, we obtain the basic evolution equations of the BH and the CFs in terms of 
four parameters related to our model, i.e., $a_*$, $\theta _M$, $k$ and $n$. We find that the evolution characteristics 
of the Kerr BH can be well described by using the CHs and the RPs in the corresponding parameter space.

It turns out that the main difference between the BZ process and MC process lies in the two kinds of loads: the remote 
load with an unknown parameter $k$ for the BZ process, and the load disc with the parameter $\beta$ for MC process. 
Parameter $k$ is uncertain, while $\beta$ is thoroughly determined by the BH spin and the place where the magnetic flux penetrates. The effects of MC process 
on the evolution of the BH and the efficiency of BH accretion disc can be discussed in virtue of the parameters $a_*$, 
$\theta _M$ and $n$, into which the parameter $\beta$ is merged.

In this model the mapping relation between the BH horizon and the disc is derived by assuming a power-law of the magnetic 
field varying as the radial coordinate of the disc. It is shown that the power-law index $n$ is related to the outer boundary 
parameter $\xi _{out}$. And the MC correction to the accretion rate is considered by using the conservation law of angular 
momentum. Only the accretion rate at the inner edge of the disc being involved in the evolution equations of the BH, the 
correction to the accretion rate is made by taking the corresponding values at $r_{ms}$ in our simplified model. 

In summary this model provides an analytic approach to the evolution characteristics and related physical quantities of a 
Kerr black hole surrounded by magnetized accretion disc.
}

\ \\ \\
{\bf ACKNOWLEDGEMENTS}
\vspace{0.2cm}
\\This work is supported by the National Natural Science Foundation of China under Grant No. 10173004. We are very 
grateful to the anonymous referee for his suggestions about the MC effects on the accretion rate and the constraints to 
the parameters of MC process.

\appendix
\section{Derivation of equation(43)}
{
In the case without magnetic coupling between the BH and the disc, accretion is produced
by the internal viscous torque $ T_{vis} $ of the disc. By using the conservation law of angular 
momentum the accretion rate $ \dot M_D $ is related to $ T_{vis} $ by the following equation (Frank, 
King \& Raine 1992):
\begin{equation}
-\dot M_D \partial ( r^2 \Omega_D) /\partial r=\partial T_{vis}/\partial r
\end{equation}
where $ \partial T_{vis}/\partial r $ is the contribution due to internal viscous torque, always 
transporting angular momentum outward in the disc. Taking MC effects into account, we think 
equation (A1) should be modified by
\begin{equation}
-\dot M_D \partial ( r^2 \Omega_D)/\partial r=\partial T_{vis}/\partial r - \partial T_{MC}/\partial r
\end{equation}
where $ \partial T_{MC}/\partial r $ is the contribution due to MC torque, and the minus sign arises
from the consideration that the contribution of $ \partial T_{vis} /\partial r $ should be counteracted
by $ \partial T_{MC}/\partial r $ in case $\Omega_D < \Omega_H $.  Comparing equations (A1) and
(A2), we express $ (\dot M_D)_{mc} $, the contribution of MC ef\/fects to the accretion rate,
as follows:
\begin{equation}
(\dot M_D)_{mc}=\frac{\partial T_{MC}}{\partial (r^2 \Omega_D)}= 
\frac {\partial T_{MC}/\partial r}{\partial (r^2 \Omega_D)/\partial r}
\end{equation}
Substituting  equations (6), (10), (12), (14) and (34) into equation (A3), we have
\begin{eqnarray}
(\dot M_D)_{mc} & = & \frac{4 {B^2}_H M^3 a_* (1+q) (1-\beta) \sin^2 \theta }{ 2- (1-q) \sin^2 \theta} \nonumber \\
    & & \times  \frac{2 (\chi^3+a_*)^2}{r_{ms} \chi^2 (\chi^3+4 a_*)} G(a_*;\xi,n)
\end{eqnarray}
From equation (A4) we find that the MC correction depends on the place on the disc where
the MC torque acts. Substituting the boundary values such as $ \xi =\xi_{in} =1 $ (i.e.,$r=r_{ms}
=M\chi^2_{ms}$) and $ \theta=\pi / 2 $ into equation (A4) we obtain equation (43), which is the 
MC correction to the accretion rate at the inner edge of the disc. In derivation the factor 2 is given for 
the two faces of the disc, and $ B^2_H $ is also taken as the average value over the BH horizon.
}


\label{lastpage}

\begin{thebibliography}{99}
\bibitem{BPT72}{Bardeen J. M., Press W. H., and Teukolsky S. A., 1972, ApJ, 178, 347}
\bibitem{Blandford76}{Blandford R. D., 1976, MNRAS, 176, 465}
\bibitem{Blandford77}{Blandford R. D., Znajek R. L., 1977, MNRAS, 179, 433}
\bibitem{Blandford99}{Blandford R. D., 1999 in {\sl Astrophysical Discs:} An EC Summer School, Astronomical 
                  Society of the Pacif\/ic Confrence Series, ed. Scllwood J A \& Goodman J 160, 265. 
                  1999 preprint  astro-ph/9902001}
\bibitem{FKR85}{Frank J., King A. R., Raine D. L., 1985, {\sl Accretion Power in Astrophysics}, 
             Cambridge Univ. Press, Cambridge}
\bibitem{GA97}{Ghosh P., Abramowicz M. A., 1997, MNRAS 292, 887 }
\bibitem{LWB00}{Lee H. K., Wijers R. A. M. J., Brown G. E., 2000, Phys. Rep., 325, 83 (LWB)
             preprint astro-ph/9906213}
\bibitem{LiLX2ka}{Li L. -X. 2000a, ApJ, 533, 115L.}
\bibitem{LiLX2kb}{Li L. -X. 2000b, preprint astro-ph/0012469}
\bibitem{LiLX2kc}{Li L. -X., Paczynski B., 2000, ApJ, 534, 197L}
\bibitem{LZY96}{Lu Y. J., Zhou Y. Y., Yu K. N., Young, E. C. M., 1996, ApJ, 472, 564}
\bibitem{MT82}{Macdonald D., Thorne K. S., 1982, MNRAS, 198, 345 (MT82)}
\bibitem{MS96}{Moderski R., Sikora M., 1996, MNRAS, 283, 854}
\bibitem{MSL97}{Moderski R., Sikora M., Lasota J.P., 1997, in {\sl "Relativistic Jets in AGNs"} eds.M. Ostrowski, 
             M. Sikora, G. Madejski \& M. Belgelman, Krakow 1997, 110, preprint astro-ph/9706263}
\bibitem{PV88}{Park S. J., Vishniac E. T., 1988, ApJ, 332, 135}
\bibitem{PC90}{Punsly B., Coroniti F. V., 1990, ApJ, 350, 518}
\bibitem{Rees84}{Rees M. J., 1984, ARA\&A. 22, 471}
\bibitem{ST83}{Shapiro S. L., Teukolsky S. A., 1983, {\sl Black Holes, White Dwarfs and Neutron Stars }, 
             John Wiley \& Sons, Inc. New York}
\bibitem{TPM86}{Thorne K. S., Price R. H., Macdonald D. A., 1986, {\sl Black Holes: The Membrane Paradigm }, 
             Yale Univ. Press, New Haven and London (TPM)}
\bibitem{Wang98}{Wang D. X., Lu Y., Yang L. T., 1998, MNRAS, 294, 667 (WLY)}
\end{thebibliography}
\end{document}